\documentclass{article}
\usepackage{graphicx}
\usepackage{tabularx}

\input{tcilatex}

\begin{document}

\author{Salman Khan\thanks{%
sksafi@phys.qau.edu.pk} and M. K. Khan \\
%EndAName
Department of Physics, Quaid-i-Azam University, \\
Islamabad 45320, Pakistan}
\title{Open Quantum Systems in Noninertial Frames}
\maketitle

\begin{abstract}
We study the effects of decoherence on the entanglement generated by Unruh
effect in noninertial frames by using bit flip, phase damping and
depolarizing channels. It is shown that decoherence strongly influences the
initial state entanglement. The entanglement sudden death can happens
irrespective of the acceleration of the noninertial frame under the action
of phase flip and phase damping channels. It is investigated that an early
sudden death happens for large acceleration under the depolarizing
environment. Moreover, the entanglement increases for a highly decohered
phase flip channel.\newline

\end{abstract}

\section{Introduction}

Entanglement is one of the potential sources of quantum theory. It is the
key concept and major resource for quantum communication and computation 
\cite{springer}. In the last few years, enormous efforts has been made to
investigate various aspects of quantum entanglement and its benefits in a
number of setups, such as teleportation of unknown states \cite{Bennett} ,
quantum key distribution \cite{Ekert}, quantum cryptography \cite{Bennett2}
and quantum computation \cite{Grover, Vincenzo}. Recently, the study of
quantum entanglement of various fields has been extended to the relativistic
setup \cite{Alsing,Ling,Gingrich,Pan, Schuller, Terashima} and interesting
results about the behavior of entanglement have been obtained. The study of
entanglement in the relativistic framework is important not only from
quantum information perspective but also to understand deeply the black hole
thermodynamics \cite{Bombelli, Callen} and the black hole information
paradox \cite{Hawking, Terashima2}.

The earlier investigations on quantum entanglement in the relativistic
framework is mainly focused by considering isolated quantum systems. In
fact, no quantum system can be completely isolated from its environment and
may results in a non-unitary dynamics of the system. Therefore, it is
important to study the effect of environment on the entanglement in an
initial state of a quantum system during its evolution. The interaction
between an environment and a quantum system leads to the phenomenon of
decoherence and it gives rise to an irreversible transfer of information
from the system to the environment \cite{Zurik, Breuer, Zurik2}. 
\begin{figure}[h]
\begin{center}
\begin{tabular}{ccc}
\vspace{-0.5cm} \includegraphics[scale=1.2]{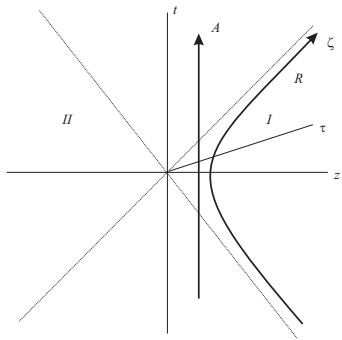}\put(-350,220) &  & 
\end{tabular}%
\end{center}
\caption{Rindler spacetime diagram: A uniformly accelerated observer Rob (R)
moves on a hyperbola \ with acceleration $a$ in region $I$ and is causally
disconnected from region $II$.}
\label{Figure1}
\end{figure}

In this paper we work out the effect of decoherence on the entanglement of
Dirac field in a noninertial system. Alsing \textit{et al} \cite{Alsing}
have shown that the entanglement between two modes of a free Dirac field is
degraded by the Unruh effect and asymptotically reaches a nonvanishing
minimum value in the infinite acceleration. We investigate that how the loss
of entanglement through Unruh effect is influenced in the presence of
decoherence by using a phase flip, a phase damping and a depolarizing
channel in the Kraus operators formalism. The effect of amplitude damping
channel on Dirac field in a noninertial system is recently studied by Wang
and Jing \cite{Wang}. We consider two observers, Alice and Rob, that share a
maximally entangled initial state of two qubits at a point in flat Minkowski
spacetime. Then Rob moves with a uniform acceleration and Alice stays
stationary. To achieve our goal, we consider two cases. In one instance we
allow only Rob's qubit to interact with a noisy environment and in the
second instance both qubits of the two observers interact with a noisy
environment. Let the two modes of Minkowski spacetime that correspond to
Alice and Rob are, respectively, given by $|n\rangle _{A}$ and $|n\rangle
_{R}$. Moreover, we assume that the observers are equipped with detectors
that are sensitive only to their respective modes and share the following
maximally entangled initial state%
\begin{equation}
|\psi \rangle _{A,R}=\frac{1}{\sqrt{2}}\left( |00\rangle _{A,R}+|11\rangle
_{A,R}\right) ,  \label{1}
\end{equation}%
where the first entry in each ket corresponds to Alice and the second entry
corresponds to Rob. From the accelerated Rob's frame, the Minkowski vacuum
state is found to be a two-mode squeezed state \cite{Alsing},%
\begin{equation}
|0\rangle _{M}=\cos r|0\rangle _{I}|0\rangle _{II}+\sin r|1\rangle
_{I}|1\rangle _{II},  \label{2}
\end{equation}%
where $\cos r=\left( e^{-2\pi \omega c/a}+1\right) ^{-1/2}$. The constant $%
\omega $, $c$ and $a$, in the exponential stand, respectively, for Dirac
particle's frequency, light's speed in vacuum and Rob's acceleration. In Eq.
(\ref{2}) the subscripts $I$ and $II$ of the kets represent the Rindler
modes in region $I$ and $II$, respectively, in the Rindler spacetime diagram
(see Fig. (\ref{Figure1})). The excited state in Minkowski spacetime is
related to Rindler modes as follow \cite{Alsing}%
\begin{equation}
|1\rangle _{M}=|1\rangle _{I}|0\rangle _{II}.  \label{3}
\end{equation}

In terms of Minkowski modes for Alice and Rindler modes for Rob, the
maximally entangled initial state of Eq. (\ref{1}) by using Eqs. (\ref{2})
and (\ref{3}) becomes%
\begin{equation}
|\psi \rangle _{A,I,II}=\frac{1}{\sqrt{2}}\left( \cos r|0\rangle
_{A}|0\rangle _{I}|0\rangle _{II}+\sin r|0\rangle _{A}|1\rangle
_{I}|1\rangle _{II}+|1\rangle _{A}|1\rangle _{I}|0\rangle _{II}\right) .
\label{4}
\end{equation}%
Since Rob is causally disconnected from region $II$, we must take trace over
all the modes in region $II$. This leaves the following mixed density matrix
between Alice and Rob, that is,%
\begin{eqnarray}
\rho _{A,I} &=&\frac{1}{2}[\cos ^{2}r|00\rangle _{A,I}\langle 00|+\cos
r(|00\rangle _{A,I}\langle 11|+|11\rangle _{A,I}\langle 00|)  \nonumber \\
&&\sin ^{2}r|01\rangle _{A,I}\langle 01|+|11\rangle _{A,I}\langle 11|].
\label{5}
\end{eqnarray}

%TCIMACRO{\TeXButton{B}{\begin{table*}[htb]}}%
%BeginExpansion
\begin{table*}[htb]%
%EndExpansion
\caption{A single qubit Kraus operators for phase flip channel, phase
damping channel and depolarizing channel. \label{table:1}}%
\begin{tabular}{|c|c|}
\hline
phase flip channel & $E_{o}=\sqrt{1-p}\left( 
\begin{array}{cc}
1 & 0 \\ 
0 & 1%
\end{array}%
\right) ,\qquad E_{1}=\sqrt{p}\left( 
\begin{array}{cc}
1 & 0 \\ 
0 & -1%
\end{array}%
\right) $ \\ \hline
phase damping channel & $E_{o}=\left( 
\begin{array}{cc}
1 & 0 \\ 
0 & \sqrt{1-p}%
\end{array}%
\right) ,\qquad E_{1}=\left( 
\begin{array}{cc}
0 & 0 \\ 
0 & \sqrt{p}%
\end{array}%
\right) $ \\ \hline
depolarizing channel & $%
\begin{array}{c}
E_{o}=\sqrt{1-p}\left( 
\begin{array}{cc}
1 & 0 \\ 
0 & 1%
\end{array}%
\right) ,\qquad E_{1}=\sqrt{p/3}\left( 
\begin{array}{cc}
0 & 1 \\ 
1 & 0%
\end{array}%
\right) , \\ 
E_{2}=\sqrt{p/3}\left( 
\begin{array}{cc}
0 & -i \\ 
i & 0%
\end{array}%
\right) ,\qquad E_{3}=\sqrt{p/3}\left( 
\begin{array}{cc}
1 & 0 \\ 
0 & -1%
\end{array}%
\right)%
\end{array}%
$ \\ \hline
\end{tabular}

%TCIMACRO{\TeXButton{End}{\end{table*}[htb]}}%
%BeginExpansion
\end{table*}[htb]%
%EndExpansion

\section{single qubit in a noisy environment}

In this section we consider that only the Rob's qubit is coupled to a noisy
environment. The final density matrix of the system in the Kraus operators
representation becomes%
\begin{equation}
\rho _{f}=\sum_{i}\left( \sigma _{o}\otimes E_{i}\right) \rho _{A,I}\left(
\sigma _{o}\otimes E_{i}^{\dag }\right) ,  \label{6}
\end{equation}%
where $\rho _{A,I}$ is the initial density matrix of the system given by Eq.
(\ref{5}), $\sigma _{o}$ is the single qubit identity matrix and $E_{i}$ are
a single qubit Kraus operators of the channel under consideration. The Kraus
operators of the channels we use are given in Table $1$. The spin-flip
matrix of the final density matrix of Eq. (\ref{6}) is defined as $\tilde{%
\rho}_{f}=\left( \sigma _{2}\otimes \sigma _{2}\right) \rho _{f}\left(
\sigma _{2}\otimes \sigma _{2}\right) $, where $\sigma _{2}$ is the Pauli
matrix. The degree of entanglement in the two qubits mixed state in a noisy
environment can be quantified conveniently by concurrence $C$, which is
given as \cite{Wootter,Coffman}%
\begin{equation}
C=\max \left\{ 0,\sqrt{\lambda _{1}}-\sqrt{\lambda _{2}}-\sqrt{\lambda _{3}}-%
\sqrt{\lambda _{4}}\right\} \qquad \lambda _{i}\geq \lambda _{i+1}\geq 0,
\label{7}
\end{equation}%
where $\lambda _{i}$ are the eigenvalues of the matrix $\rho _{f}\tilde{\rho}%
_{f}$. The eigenvalues under the action of phase-flip channel becomes

\begin{figure}[h]
\begin{center}
\begin{tabular}{ccc}
\vspace{-0.5cm} \includegraphics[scale=1.2]{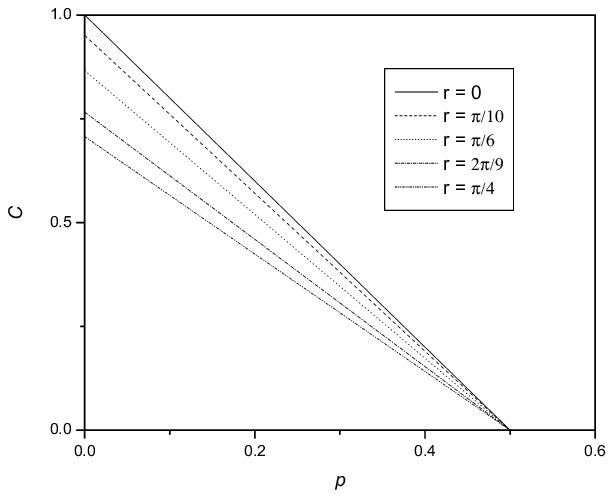}\put(-350,220) &  & 
\end{tabular}%
\end{center}
\caption{The concurrence $C$ under the action of phase flip channel is
plotted against decoherence parameter $p$ for the case when only Rob's qubit
is coupled to a noisy environment.}
\label{Figure2}
\end{figure}
\begin{eqnarray}
\lambda _{1}^{\mathrm{PF}} &=&(1-2p+p^{2})\cos ^{2}r,  \nonumber \\
\lambda _{2}^{\mathrm{PF}} &=&p^{2}\cos ^{2}r,  \nonumber \\
\lambda _{3}^{\mathrm{PF}} &=&\lambda _{4}^{\mathrm{PF}}=0,  \label{8}
\end{eqnarray}%
where the superscript \textrm{PF} corresponds to phase flip channel.
Similarly, the eigenvalues under the action of phase damping and
depolarizing channels are, respectively, given by%
\begin{eqnarray}
\lambda _{1,2}^{\mathrm{PD}} &=&\frac{1}{4}(2-p\pm 2\sqrt{1-p})\cos ^{2}r, 
\nonumber \\
\lambda _{3}^{\mathrm{PD}} &=&\lambda _{4}^{\mathrm{PD}}=0,  \label{10}
\end{eqnarray}%
\begin{eqnarray}
\lambda _{1}^{\mathrm{DP}} &=&(-1+p)^{2}\cos ^{2}r,  \nonumber \\
\lambda _{2}^{\mathrm{DP}} &=&\lambda _{3}^{\mathrm{DP}}=\lambda _{4}^{%
\mathrm{DP}}=\frac{1}{9}p^{2}\cos ^{2}r,  \label{11}
\end{eqnarray}%
where the superscripts \textrm{PD} and \textrm{DP} stand for phase damping
and depolarizing channels, respectively. In all these equations $p\in \left[
0,1\right] $ is the decoherence parameter. The upper and lower values of $p$
correspond to undecohered and fully decohered case of the channels,
respectively. The concurrence under the action of every channel reduces to
the result of Ref. \cite{Alsing} when the decoherence parameter $p=0$.

To see how the concurrence and hence the entanglement is influenced by
decoherence parameter $p$ in the presence of Unruh effect, we plot the
concurrence for each channel against $p$ for various values of $r$. In Fig. (%
\ref{Figure2}), the concurrence under the action of phase flip channel is
plotted against $p$. The figure shows that for smaller values of $p$, the
entanglement is strongly acceleration dependent, such that for large values
of Rob's acceleration (the value of $r$) it gets weakened. However, as $p$
increases the dependence of entanglement on acceleration decreases and the
increasing value of $p$ causes a rapid loss of entanglement. 
\begin{figure}[h]
\begin{center}
\begin{tabular}{ccc}
\vspace{-0.5cm} \includegraphics[scale=1.2]{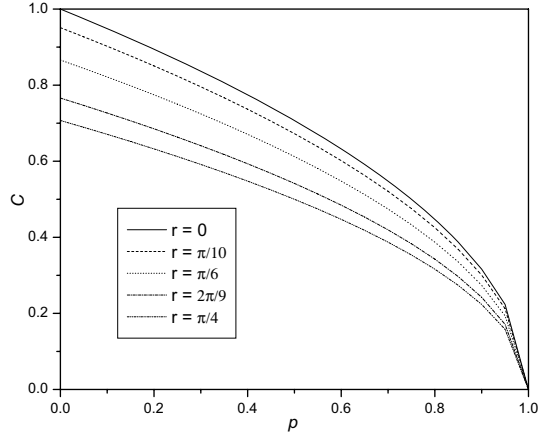}\put(-350,220) &  & 
\end{tabular}%
\end{center}
\caption{The concurrence $C$ under the action of phase damping channel is
plotted against decoherence parameter $p$ for the case when only Rob's qubit
is coupled to a noisy environment.}
\label{Fig3}
\end{figure}
\ The entanglement sudden death happens irrespective of the acceleration of
Rob's frame for a $50\%$ decoherence. Fig. (\ref{Fig3}) shows the effect of
decoherence on the concurrence under the action of phase damping channel. In
this case, the degradation of entanglement due to decoherence is smaller as
compare to the the degradation in the case of phase flip. The entanglement
vanishes for all values of acceleration only when the channel is fully
decohered. The concurrence under the action of the depolarizing channel is
exactly equal to the one for phase flip channel. Hence it influences the
entanglement in a way exactly similar to the phase flip channel as shown in
Fig. ($2$).

\section{Both qubits in a noisy environment}

In this section we consider that both Alice's and Rob's qubits are
influenced simultaneously by a noisy environment. The final density matrix
in this case can be written in the Kraus operators formalism as follows%
\begin{equation}
\rho _{f}=\sum_{k}E_{k}\rho _{A,I}E_{k}^{\dag },  \label{12}
\end{equation}%
where $\rho _{A,I}$ is given by Eq. (\ref{5}) and $E_{k}$ are the Kraus
operators for a two qubit system, satisfying the completeness relation $%
\sum_{k}E_{k}E_{k}=I$ and are constructed from a single qubit Kraus
operators of a channel by taking tensor product of all the possible
combinations in the following way%
\begin{equation}
E_{k}=\sum_{i,j}E_{i}\otimes E_{j},  \label{13}
\end{equation}%
where $E_{i,j}$ are the single qubit Kraus operators of a channel given in
Table $1$. We consider that both Alice's and Bob's qubits are influenced by
the same environment, that is, the decoherence parameter $p$ for both qubits
is same. Proceeding in a similar way like the case of single qubit coupled
to the environment, the eigenvalues of the matrix $\rho _{f}\tilde{\rho}_{f}$
under the action of phase flip channel become

\begin{eqnarray}
\lambda _{1}^{\mathrm{PF}} &=&(1+2(-1+p)p)^{2}\cos ^{2}r,  \nonumber \\
\lambda _{2}^{\mathrm{PF}} &=&4(-1+p)^{2}p^{2}\cos ^{2}r,  \nonumber \\
\lambda _{3}^{\mathrm{PF}} &=&\lambda _{4}^{\mathrm{PF}}=0,  \label{14}
\end{eqnarray}%
Likewise the eigenvalues for phase damping and depolarizing channels,
respectively, becomes

\begin{eqnarray}
\lambda _{1}^{\mathrm{PD}} &=&\frac{1}{4}(-2+p)^{2}\cos ^{2}r,  \nonumber \\
\lambda _{2}^{\mathrm{PD}} &=&\frac{1}{4}p^{2}\cos ^{2}r,  \nonumber \\
\lambda _{3}^{\mathrm{PD}} &=&\lambda _{4}^{\mathrm{PD}}=0,  \label{16}
\end{eqnarray}

\begin{eqnarray}
\lambda _{1,3}^{\mathrm{DP}} &=&\frac{1}{1296}[324+p(-3+2p)(387+152p(-3+2p))
\nonumber \\
&&+4(3-4p)^{2}(9+5p(-3+2p))\cos 2r  \nonumber \\
&&+(3-4p)^{2}p(-3+2p)\cos 4r\pm 4(3-4p)^{2}\cos r  \nonumber \\
&&\times \{3(54+p(-3+2p)(33+8p(-3+2p)))  \nonumber \\
&&+(3-4p)^{2}(2(9-6p+4p^{2})\cos 2r+p(-3+2p)\cos 4r)\}^{1/2}],  \nonumber \\
\lambda _{2}^{\mathrm{DP}} &=&\lambda _{4}^{\mathrm{DP}}=\frac{1}{648}%
p(-3+2p)(-9+4p  \nonumber \\
&&+(-3+4p)\cos 2r)(3+4p+(-3+4p)\cos 2r),  \label{17}
\end{eqnarray}%
The $"\pm "$ sign in Eq. (\ref{17}), correspond to the eigenvalues $\lambda
_{1}$, and $\lambda _{3}$ respectively. It is necessary to point out here
that the concurrence under the action of each channel reduces to the result
of Ref. \cite{Alsing} when we set the decoherence parameter $p=0$. 
\begin{figure}[h]
\begin{center}
\begin{tabular}{ccc}
\vspace{-0.5cm} \includegraphics[scale=1.2]{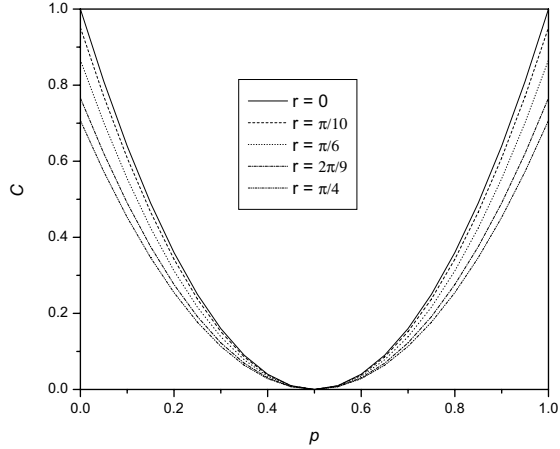}\put(-350,220) &  & 
\end{tabular}%
\end{center}
\caption{The concurrence $C$ under the action of phase flip channel is
plotted against decoherence parameter $p$ for the case when both qubits are
coupled to a noisy environment.}
\label{Fig4}
\end{figure}

To see how the entanglement behaves when both the qubits are coupled to the
noisy environment, we plot the concurrence against the decoherence parameter 
$p$ for different values of $r$ under the action of each channel separately.
Fig. (\ref{Fig4}) shows the dependence of concurrence on decoherence
parameter $p$ under the action of phase flip channel. The dependence of
entanglement on acceleration of Rob's frame is obvious in the region of
lower values of $p$. However, this dependence diminishes as $p$ increases
and a rapid decrease in the degree of entanglement develops. At a $50\%$
decoherence level, the entanglement sudden death occurs irrespective of
Rob's acceleration. It's interesting to see that beyond this point onward,
the entanglement regrows as $p$ increases. The dependence of entanglement on
acceleration of the Rob's frame reemerges and the entanglement reaches to
the corresponding undecohered maximum value for a fully decohered case. The
concurrence varies as a parabolic function of decoherence parameter $p$ with
its vertex at $p=0.5$. 
\begin{figure}[h]
\begin{center}
\begin{tabular}{ccc}
\vspace{-0.5cm} \includegraphics[scale=1.2]{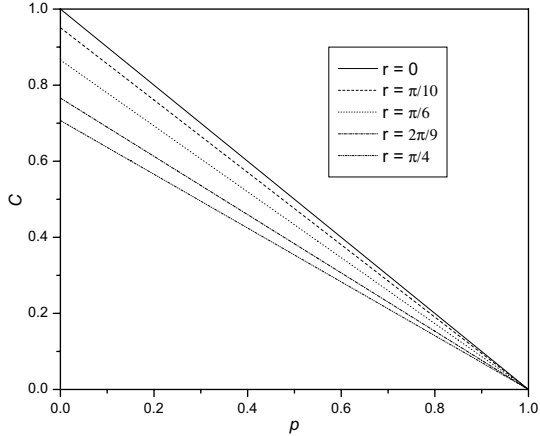}\put(-350,220) &  & 
\end{tabular}%
\end{center}
\caption{The concurrence $C$ under the action of phase damping channel is
plotted against decoherence parameter $p$ for the case when both qubits are
coupled to a noisy environment.}
\label{Fig5}
\end{figure}
The dependence of entanglement on $p$ under the action of phase damping
channel is shown in Fig. (\ref{Fig5}). In this case the entanglement
decreases linearly as $p$ increases and the dependence on acceleration
diminishes. Whatever the acceleration of Rob's frame may be, the
entanglement sudden death occurs when the channel is fully decohered. The
influence of depolarizing channel on the entanglement is shown in Fig. (\ref%
{Fig6}). Unlike the other two channels, the depolarizing channel does not
diminish the effect of acceleration on the entanglement as the $p$
increases. However a rapid decrease in entanglement appears which leads to
entanglement sudden death at different values of decoherence parameter for
different acceleration of Rob's frame. The larger the acceleration the
earlier the entanglement sudden death occurs. 
\begin{figure}[h]
\begin{center}
\begin{tabular}{ccc}
\vspace{-0.5cm} \includegraphics[scale=1.2]{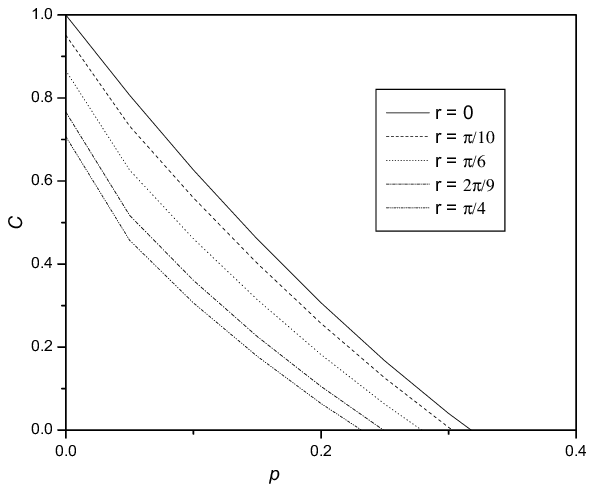}\put(-350,220) &  & 
\end{tabular}%
\end{center}
\caption{The concurrence $C$ under the action of depolarizing channel is
plotted against decoherence parameter $p$ for the case when both qubits are
coupled to a noisy environment.}
\label{Fig6}
\end{figure}

If we compare the single qubit and the both qubits decohering situations, it
becomes obvious that the entanglement loss is rapid when both the qubits are
coupled to the noisy environment. For example, in the case of bit flip
channel the concurrence behaves as a linear function of $p$ for single qubit
decohering case whereas in the case of both qubits decohering case it varies
as a parabolic function. Nevertheless, the sudden death happens at the same
value of $p$, irrespective of the acceleration, for both cases under the
action of bit flip and phase damping channels. For depolarizing channel,
however, this is not true.

\section{Conclusion}

In conclusion, we have investigated that the entanglement in Dirac fields is
strongly dependent on coupling with a noisy environment. This result is
contrary to the case of an isolated system in which the entanglement of
Dirac fields survives even in the limit of infinite acceleration of Rob's
frame. In the presence of decoherence, the entanglement rapidly decreases
and entanglement sudden death occurs even for zero acceleration. Under the
action of phase flip channel, the entanglement can regrow when both qubits
are coupled to a noisy environment in the limit of large values of
decoherence parameter. The entanglement disappears, irrespective of the
acceleration, under the action of phase damping channel only when the
channel is fully decohered both for single qubit and the two qubits
decohering cases. However, under the action of depolarizing channel an early
sudden death occurs for larger acceleration when both qubits are coupled to
the environment. In summary, the entanglement generated by Unruh effect in
noninertial frame is strongly influenced by decoherence.

\end{document}